# RF CONTROLLABLE IOFFE-PRITCHARD TRAP

GUENNADI A. KOUZAEV

*Department of Electronics and Telecommunications, Norwegian University of Science and Technology - NTNU, Trondheim, N-7491 Norway*
*guennadi.kouzaev@iet.ntnu.no*

KARL J. SAND

*Department of Electronics and Telecommunications, Norwegian University of Science and Technology - NTNU, Trondheim, N-7491 Norway*
*sand@iet.ntnu.no*

An Ioffe-Pritchard trap for cold dressed atoms is studied by analytical and numerical simulations. The effective potential in this trap is formed by the static magnetic and radio-frequency fields, and the minimums are formed around the current bars. The depth of the minimums and the overall topology of the effective potential are controlled electronically. The studied regime of the Ioffe-Pritchard trap is of interest for high-sensitive cold atom interferometers. Submitted to Modern Physics Letters B 31.01.2006.

1. Introduction

For a long period of time the magnetic traps have been an effective mean to study quantum physics of clouds of cold atoms[1-4]. The atoms are confined inside the three-dimensional traps or close to the surface of a microchip with lithographically placed planar conductors.[5,6] Traps are installed inside a vacuum chamber, and laser light is used to illuminate the atomic clouds to study the spatial distribution of atoms.

   Recent publications show that the magnetic traps are able not only to confine atoms but also of realizing important manipulation of atom clouds such as the splitting of the atom clouds by a specially formed static magnetic field along the surface of the chip.[4] A special attention has been paid to the traps initiated by static (DC) magnetic and radio-frequency (RF) fields. For example, the RF-field rotating around the longitudinal *z*-axis is used for additional pinching of the atom clouds or to control the distance between the minimums of the double-well potential.[7,8]

New results are derived for the traps handling dressed atoms by a combination of the static magnetic and RF fields. In this case, these atoms interact with a large number of electromagnetic quanta, and new topologies of the trapping potential can be realized.[2,8-12] Recently, this approach was applied for frequency controllable splitting of the atom clouds traveling along a wire carrying a static current.[11] The RF and DC fields form a double-well potential along the wire, and the transverse distance between the minima of the effective adiabatic potential is controlled by the RF frequency. A ring-like interferometer is considered in[12]. These published experiments and models show the possibility to realize quantum interferometers and quantum gates based on the dressed atom formalism.

In this paper, we study the effective potentials for cold dressed atoms in the RF controllable Ioffe-Pritchard trap (Fig. 1). It is found that the formation of the multi-minimum potential is controllable by the RF frequency and DC trap currents. The derived results are interesting in the study of the interferometric quantum effects for cold atoms.

2.  Analytical Modeling of the Effective Potential for Cold Dressed Atoms

The effective potential $U_{\text{eff}}(\mathbf{r})$ induced by a combination of the RF and constant fields is computed according to a formula [11]

$$U_{\text{eff}}(\mathbf{r}) = m_F \sqrt{\left[\mu_B g_F |\mathbf{B}_{\text{DC}}(\mathbf{r})| - \hbar\omega_{\text{RF}}\right]^2 + \left[\mu_B g_F B_{\text{RF}}^{(\perp)}(\mathbf{r})/2\right]^2} \quad (1)$$

where $m_F = 2$ is the magnetic quantum number of the atomic state, $\mu_B$ is the Bohr magneton, $g_F$ is the Landè factor, $\mathbf{B}_{\text{DC}}(\mathbf{r})$ is the static magnetic trap field, $\hbar$ is the reduced Planck constant, $\omega_{RF}$ is the cyclic frequency of the RF field, and $B_{\text{RF}}^{(\perp)}$ is the RF magnetic field normal to the local DC magnetic vector

$$B_{\text{RF}}^{(\perp)}(\mathbf{r}) = \sqrt{\left(\mathbf{B}_{\text{RF}}(\mathbf{r})\right)^2 - \left(\frac{\mathbf{B}_{\text{RF}}(\mathbf{r})\mathbf{B}_{\text{DC}}(\mathbf{r})}{|\mathbf{B}_{\text{DC}}(\mathbf{r})|}\right)^2}. \quad (2)$$

The potential $U_{\text{eff}}(\mathbf{r})$ is the parametrically depending function on the RF frequency $\omega_{\text{RF}}$ and the geometry of the trap. It has a local minimum where $\mu_B g_F |\mathbf{B}_{\text{DC}}(\mathbf{r})| = \hbar\omega_{\text{RF}}$. Depending on the spatial distribution of the static magnetic field, the minimum can be like an egg-shell or a line.[9,10] More complicated trap geometries can provide potentials with the controllable minimum depth. The studied Ioffe-Pritchard trap consists of four bars and two rings (Fig.1a,b), and the topology of the effective potential varies with the geometry of this trap, currents, and the frequency.

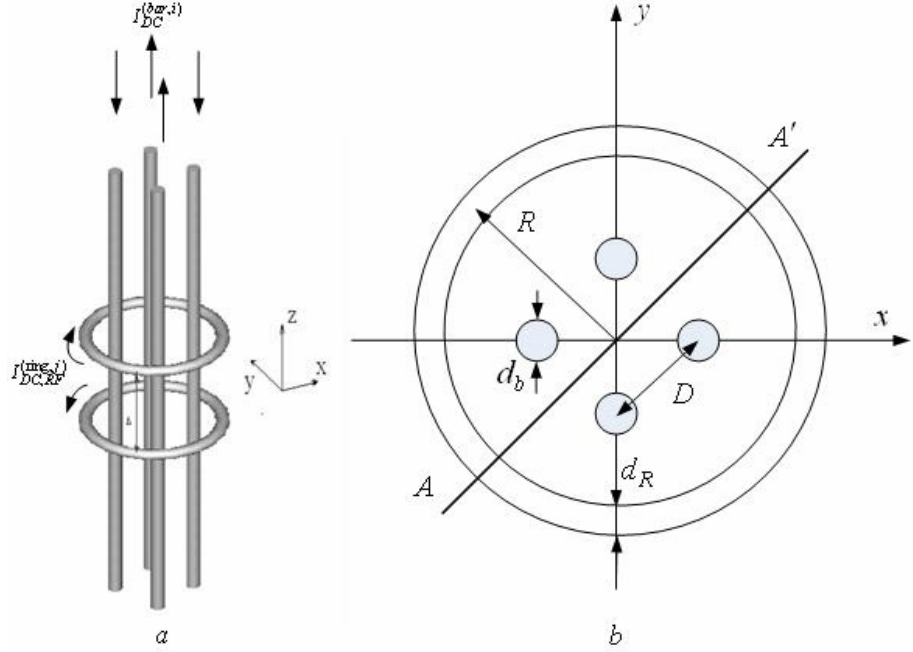

Fig. 1. 3D-view of the Ioffe-Pritchard trap (a) and its cross-section (b).

3.     RF Induced Potential in the Ioffe-Pritchard Trap

The geometry of the Ioffe-Pritchard trap is shown in Fig. 1a,b. Four bars support the static currents. Two rings carry the DC currents in opposite directions to each other. In addition to the DC currents, the rings excite the RF field due to alternating currents.

The static magnetic field $\mathbf{B}_{DC}(\mathbf{r})$ in the trap is computed as the absolute value of the sum of the static magnetic fields of the bars $\mathbf{B}^{(bar)}$ and rings $\mathbf{B}^{(ring)}$. The first is the sum of the fields from four bars with the coordinates $(x_i, y_i,\ i=1,...,4)$ [4]

$$B^{(bar,i)}_{x_{DC}}(x,y) = -\frac{\mu_0 I^{(bar,i)}_{DC} \cdot \left(y - \Delta_i^{(y)}\right)}{2\pi r_i^2}, \tag{3}$$

$$B^{(bar,i)}_{y_{DC}}(x,y) = \frac{\mu_0 I^{(bar,i)}_{DC} \cdot \left(x - \Delta_i^{(x)}\right)}{2\pi r_i^2} \tag{4}$$

where $I^{(bar,i)}_{DC}$ is the current on the $i$th bar, $r_i = \sqrt{\left(x-\Delta_i^{(x)}\right)^2 + \left(y-\Delta_i^{(y)}\right)^2}$, $\Delta_i^{(x)}$ and $\Delta_i^{(y)}$ are the shifts of the bars from the coordinate system centre $(x=y=0)$, $d_b = 0$, and $\mu_0 = 4\pi \cdot 10^{-7}\ \mathrm{H/m}$.

The static magnetic field $\mathbf{B}^{(ring)}$ is the sum of the fields $\mathbf{B}^{(ring,j)}\left(z_{r_j}\right)$ from $j$th ring placed at $z = z_{r_j}$

$$\mathbf{B}^{(\text{ring})} = \sum_{j=1}^{2} \mathbf{B}^{(\text{ring},j)}\left(z_{r_j}\right) \qquad (5)$$

where $\mathbf{B}^{(\text{ring},j)}\left(z_{r_j}\right)$ is the magnetic field[1]

$$B_{z_{\text{DC}}}^{(\text{ring},j)}(x,y,z) = \frac{\mu_0 I_{\text{DC}}^{(\text{ring},j)}}{2\pi} \cdot \frac{1}{\left[(R+\rho)^2 + (z-z_{r_j})^2\right]^{0.5}} \cdot \left[K(k^2) + \frac{R^2 - \rho^2 - (z-z_{r_j})^2}{(R-\rho)^2 + (z-z_{r_j})^2} E(k^2)\right], \quad (6)$$

$$B_{x_{\text{DC}}}^{(\text{ring},j)}(x,y,z) = \frac{\mu_0 I_{\text{DC}}^{(\text{ring},j)} x}{2\pi \rho^2} \cdot \frac{z-z_{r_j}}{\left[(R+\rho)^2 + (z-z_{r_j})^2\right]^{0.5}} \cdot \left[-K(k^2) + \frac{R^2 + \rho^2 + (z-z_{r_j})^2}{(R-\rho)^2 + (z-z_{r_j})^2} E(k^2)\right], \quad (7)$$

$$B_{y_{\text{DC}}}^{(\text{ring},j)}(x,y,z) = \frac{\mu_0 I_{\text{DC}}^{(\text{ring},j)} y}{2\pi \rho^2} \cdot \frac{z-z_{r_j}}{\left[(R+\rho)^2 + (z-z_{r_j})^2\right]^{0.5}} \cdot \left[-K(k^2) + \frac{R^2 + \rho^2 + (z-z_{r_j})^2}{(R-\rho)^2 + (z-z_{r_j})^2} E(k^2)\right] \quad (8)$$

where $K$ and $E$ are the elliptic integrals with the argument

$$k^2 = \frac{4R\rho}{(R+\rho)^2 + (z-z_{r_j})^2}. \qquad (9)$$

In (6)-(8), $I_{\text{DC}}^{(\text{ring},j)}$ is the current on the *j*th ring, $R$ is the radius of this infinitely thin ring $(d_r = 0)$, $z_{r_j}$ is the position of the centre of the *j*th ring on the *z*-axis, and $\rho = \sqrt{x^2 + y^2}$. The overall static magnetic $\mathbf{B}_{\text{DC}}(\mathbf{r})$ and RF magnetic $\mathbf{B}_{\text{RF}}(\mathbf{r})$ fields are computed by vectorial summation of the corresponding fields from the bars and rings.

The RF field can be computed by the solution of the full Maxwell equations. Fortunately, the frequency $F$ of the RF current is below the Larmor frequency that is comparable with 1 MHz depending on the value of the static magnetic field. At such a low frequency, if the size of a trap is essentially less than a tenth of the wavelength $\lambda_0 = 3$ m, then the quasi-static approximation can be used. In this case, the RF field is equivalent to the static magnetic field.

We simulate the Ioffe-Pritchard trap shown in Fig. 1 and described above. The distance between the infinitely thin bars is $D=4$ mm, the radius of the infinitely thin rings is $R= 5$ mm, and the distance between them is $\Delta = 2|z_{r_j}| = 10$ mm. The centre of this trap is placed at the origin of the Cartesian coordinate system.

Fig. 2a shows the potential computed analytically at $z=0$ when the trap carries only DC currents. In this case, $\left|I_{\text{DC}}^{(\text{ring},j)}\right| = 0.5$ A, $j = 1, 2$ and $\left|I_{\text{DC}}^{(\text{bar},i)}\right| = 0.5$ A, $i = 1,...,4$. The potential minimum is formed at the centre between the bars. Due to the DC currents on the rings, the global minimum of the potential $U_{\text{eff}}$ is at the point $x=y=z=0$.

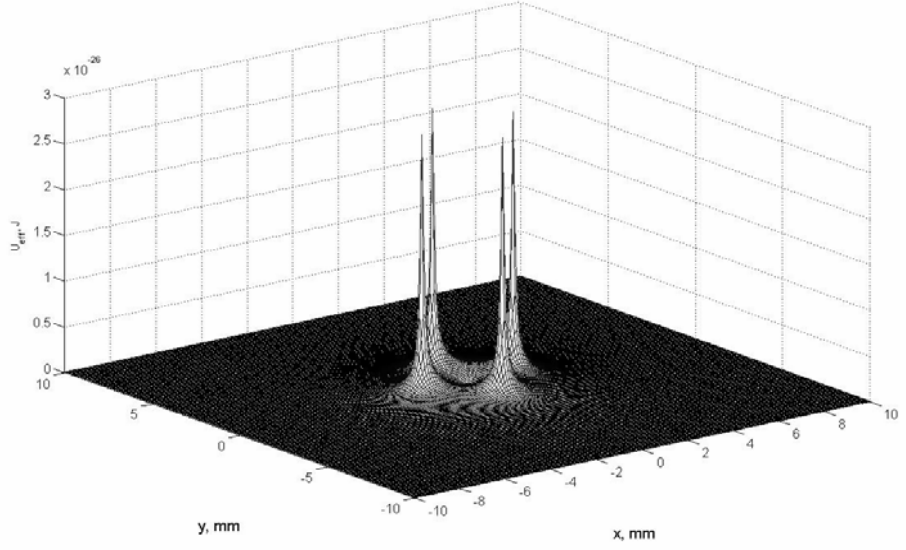

Fig. 2a. Effective potential in the Ioffe-Pritchard trap produced by the DC ring and DC bar currents ($z=0$).

The ring RF currents change the spatial topology of the effective potential. Fig. 2b, $z=0$ shows the formation of the minimums around the bars due to the dressing effect in the trap with $\left|I_{RF}^{(ring,j)}\right| = 0.1$ A, $F = 0.8$ MHz, $\left|I_{DC}^{(ring,j)}\right| = 0.5$ A, and $\left|I_{DC}^{(bar,i)}\right| = 0.1$ A at $z=0$. The RF currents of the rings have the opposite phase. At the centre of the trap with $x = y = 0$, the potential has the relative maximum with respect to the surrounding minimum. To explain this maximum origin, the effective potential formula (1) is rewritten into the following form

$$U_{\text{eff}}(\mathbf{r}) = m_F \sqrt{\left[\mu_B g_F |\mathbf{B}_{DC}(\mathbf{r})|\right]^2 + \left[(\hbar\omega_{RF})^2 + \left(\mu_B g_F B_{RF}^{(\perp)}(\mathbf{r})/2\right)^2 - 2\hbar\omega_{RF}\mu_B g_F |\mathbf{B}_{DC}(\mathbf{r})|\right]}. \qquad (10)$$

It follows that the second term in (10) can be zero under certain conditions. In this case, the local maximum of the effective potential is proportional to the value of the magnetic field at this point. Thus, under some circumstances, such an isolated maximum can attract strong-field seeking atoms with $m_F = 1$, and the trap can be loaded by two sorts of atoms localized at the centre of the trap $(m_F = 1)$ and close to the bars $(m_F = 2)$. Such a trap is interesting in research of collision of atoms with different spins for prospective quantum interferometry and quantum gates.[13]

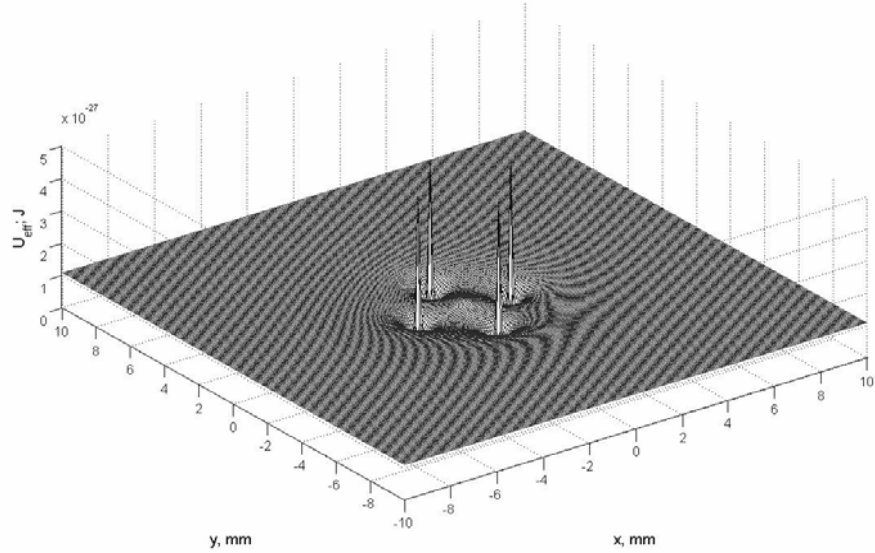

Fig. 2b. Multi-minimum effective potential at the centre of the trap ($z=0$) of the Ioffe-Pritchard trap fed by the RF and DC currents

4.   Conclusions

The Ioffe-Pritchard trap for cold dressed atoms has been studied. An effective multi-minimum potential are formed with the static magnetic and radio-frequency fields. The spatial topology of the potential is controlled by the frequency and the static currents. Changes in the currents and frequency allow the splitting of the atom clouds around the trap bars and then transforming back this topology into a single-minimum potential. The derived results are interesting in the development of new electronically controlled traps and quantum interferometers for cold atoms.